\documentclass[12pt]{article}
\usepackage[margin=25mm]{geometry}
\usepackage{amsmath}
\usepackage{graphicx,psfrag,epsf}
\usepackage{enumerate}
\usepackage{natbib}
\usepackage{amssymb}
\usepackage{epstopdf}
\usepackage{enumerate}
\usepackage{placeins}
\usepackage{setspace}
\usepackage{float}
\usepackage{soul,color}
\usepackage{placeins}
\usepackage{upgreek, mathtools,bm}
\usepackage{algorithm,algpseudocode,tabularx}
\usepackage{caption}
\usepackage{subcaption}
\usepackage{enumitem}
\usepackage{multirow}
\usepackage{amsfonts,amssymb,graphics,amsthm}
\usepackage{url}
\RequirePackage[colorlinks,breaklinks,linkcolor=blue,citecolor=blue,urlcolor=blue]{hyperref}
\usepackage{multicol}
\usepackage{listings} 
\usepackage{xcolor} 
\usepackage{booktabs}
\usepackage{bbm}
\immediate\write18{texcount -tex -sum  \jobname.tex > \jobname.wordcount.tex}

\newtheorem{assumption}{Assumption}
\newtheorem{corollary}{Corollary}

\def\p{{\bf p}}
\def\x{{\bf x}}

\def\X{{\bf X}}

\def\0{{\bf 0}}
\def\1{{\bf 1}}

\def\bb{{\boldsymbol\beta}}

\def\bbP{{\mathbb{P}}}

\def\calC{{\mathcal{C}}}
\def\calF{{\mathcal{F}}}
\def\calG{{\mathcal{G}}}

\def\calS{{\mathcal{S}}}

\def\trans{^{\rm T}}

\newcommand{\argmax}{\mathop{\mathrm{argmax}}}

\newsavebox{\coloredquotationbox}

\newcommand*\de{\mathop{}\!\mathrm{d}}
\def\minpi{{\underline{\pi}}}

\def\bbP{{\mathbb{P}}}

\graphicspath{{./Fig/}}

\newtheorem{theorem}{Theorem}[section]
\newtheorem{proposition}[theorem]{Proposition}

\theoremstyle{remark}

\theoremstyle{plain}
\theoremstyle{definition}
\newtheorem{remark}{{\bf Remark}}

\def\blue{\textcolor{black}}

\providecommand{\keywords}[1]
{
  \small	
  \textbf{\textit{Keywords---}} #1
}

\title{Modeling positive and unlabeled data with a generalized additive density ratio model}
\author{Peijun Sang$^{1}$, Yifan Sun$^{1}$, Qinglong Tian$^{1}$, Donglin Zeng$^{2}$ and Pengfei Li$^{1}$ \\
        \small $^{1}$Department of Statistics and Actuarial Science, University of Waterloo \\
        \small $^{2}$Department of Biostatistics, University of Michigan \\
}
\begin{document}
\maketitle

\begin{abstract}
This paper focuses on learning from positive and unlabeled (PU) data, where only some positives are labeled and the rest are mixed with negatives. 
Classical exponential tilting models guarantee identifiability by imposing a linear structure, but they can be severely misspecified when the true relationships are nonlinear.
We propose a generalized additive density-ratio framework that retains identifiability while allowing nonlinear and feature-specific effects. 
The approach comes with a practical fitting algorithm and supporting theory that enable estimation and inference for the mixture proportion and other quantities of interest. 
In simulations and analyses of real datasets, the proposed method matches the standard exponential tilting method when the linear model is correct and delivers clear gains when it is not. Overall, the framework strikes a useful balance between flexibility and interpretability for PU data and provides principled tools for estimation, prediction, and uncertainty assessment.
\end{abstract} \hspace{10pt}


\section{Introduction}
\label{sec:introduction}

\subsection{Positive and unlabeled data}
\label{sub:PUdata}

Consider a joint distribution over \((\X, Y)\), where \(\X\) denotes a vector of features and \(Y \in \{0, 1\}\) is a binary label.  
Positive and unlabeled (PU) data, also known as presence-only or contaminated case-control data, consist of two samples.  
The first is a labeled sample of positive instances:
$
\left\{(\x_1, y_1 = 1), \dots, (\x_{n_0}, y_{n_0} = 1)\right\}.
$
The second is an unlabeled sample containing a mixture of both positive (\(Y = 1\)) and negative (\(Y = 0\)) instances:
$
\left\{\x_{n_0 + 1}, \dots, \x_n\right\}.
$

A commonly used assumption in PU data analysis is called \emph{selected completely at random} (SCAR) (\citealt{bekker_learning_2020}).
Under SCAR, the generative process of the data is modeled as:
\begin{equation}
	\label{eq:pu-dist}
	\begin{split}
		\x_1, \dots, \x_{n_0} &\sim g(\x), \\
		\x_{n_0 + 1}, \dots, \x_n &\sim \pi g(\x) + (1 - \pi) h(\x),
	\end{split}
\end{equation}
where \(\pi \in (0,1)\) is the mixture proportion, and \(g(\x) = p(\x \mid Y = 1)\) and \(h(\x) = p(\x \mid Y = 0)\) denote the feature distributions for positive and negative classes, respectively. \blue{Thus the log-density ratio is defined as $\log\{g(\x)/h(\x)\}$.}
%For example, in a contaminated case-control design, \(\pi\) represents the fraction of undiagnosed positive cases within the contaminated control group. 
%In the context of outlier detection, \(g(\x)\) represents the distribution of typical observations, while \(h(\x)\) models the anomalous data. 

Positive and unlabeled data arise naturally in a wide range of real-world settings, particularly in biomedical research, econometrics, and machine learning \citep{jaskie_positive_2019, li2022positive}. 
We present two concrete examples drawn from the literature: one from sociology and one from ecology.
Both are based on real datasets and substantive scientific questions. 
These examples will be revisited in the numerical studies to demonstrate the empirical relevance and applicability of the proposed methodology.

%Such data arise frequently in areas including biomedical \syf{studies}, econometrics, and machine learning \citep{jaskie_positive_2019, li2022positive}. 
%For example, in case-control studies, underdiagnosis can result in true cases being mislabeled as controls, leading to a contaminated control group that includes both positive and negative instances while the case group is considered a pure sample.  
%Additional real-world examples of PU data can be found in \citet{jaskie_positive_2019}. 

\noindent\textbf{Survey data with dishonest responses}\,\,
\textit{Undeclared work}, defined as paid activities that are legal in nature but deliberately concealed from public authorities, is a central topic of the \textit{European Commission’s Special Eurobarometer Survey 402} \citep{arezzo2024estimating}. 
In this survey, only a small fraction of respondents report having engaged in undeclared work, a proportion that is likely substantially lower than the true prevalence due to misreporting or nonresponse.
Let \( Y \) denote the indicator that an individual has engaged in undeclared work. The positive-labeled sample consists of respondents who answer “yes,” whereas those who answer “no” or decline to respond form the unlabeled sample. The covariate vector \( \X \) includes demographic characteristics, financial conditions, and attitudes toward regulatory compliance. 
%Here, \( \pi \) represents the proportion of individuals who engaged in undeclared work but did not report such activity in the survey. 
Systematic underreporting in survey responses may lead to severe underestimation, thereby affecting the allocation of regulatory resources and the design of effective enforcement strategies.
Thus, accurately estimating the true prevalence of undeclared work is of considerable policy importance. 

%\noindent\textbf{Survey data with dishonest answers}\,\,
%\textit{Undeclared work}, defined as paid activities that are legal but concealed from public authorities, is a key focus of the \textit{European Commission’s Special Eurobarometer Survey 402} \citep{arezzo2024estimating}.  
%Only a small proportion of respondents admit having been engaged in undeclared work, which may be substantially lower than the true prevalence due to misreporting. Let $Y$ denote the indicator that an individual was engaged in undeclared work. The positive-labeled sample consists of respondents who answer ``yes'', while those who answer ``no'' or do not respond constitute the unlabeled sample. The covariate $\X$ summarizes demographics, financial conditions, and compliance attitudes.  
%Estimating the true size of undeclared work is important because misreporting in surveys can lead to a substantial underestimation of its prevalence, potentially hindering the design of effective policy measures.

\noindent\textbf{Species presence-only data}\,\,
\citet{elith2020presence} describe datasets containing recorded occurrence locations for numerous species across different geographic regions. 
For a given species, the positive-labeled sample consists of opportunistically collected presence records (\( Y = 1 \)). 
In addition, a set of spatially random “background” locations is selected, which forms the unlabeled sample.
Because true absences are not observed in those ``background'' locations, the background sample represents a mixture of actual presences and absences.
Environmental covariates \( \X \), such as climate and land-use variables, are available at both presence and background locations. 
%Here, \( \pi \) can be interpreted as the overall probability of species being present within the study region \citep{hastie2013inference}.
A primary objective is to characterize the relationship between species occurrence and environmental covariates using the presence records together with the background sample.

For the analysis of PU data, we are mainly interested in estimating and inferring the mixing proportion $\pi$ and the log-density ratio. Statistical inference of $\pi$ enables us to have a better idea of the proportion of positive features in the unlabeled group, and that of the density ratio indicates which features are useful to distinguish the positive samples from the negative ones.  \blue{Then we are able to predict} the label of features in the unlabeled group.
These tasks, however, are not feasible without appropriate assumptions that guarantee model identifiability.

%\vspace{-6.65pt}
%\vspace{-3.4pt}

\subsection{The dilemma between identifiability and flexibility}
\label{sub:exponential_tilting_model}

%\vspace{-3.4pt}

Without additional assumptions, the model defined in~\eqref{eq:pu-dist} is not identifiable.  
To see this, consider that for any given pair \((\pi, h(\x))\), one can construct a different pair \((\pi^*, h^*(\x))\) such that
%\begin{equation*}
$\pi g(\x) + (1 - \pi) h(\x) = \pi^* g(\x) + (1 - \pi^*) h^*(\x),$
%\end{equation*}
for any \(\pi^* \in (0, \pi)\), where
\begin{equation}
	\label{eq:not-identifiable}
	h^*(\x) = \frac{\pi - \pi^*}{1 - \pi^*} g(\x) + \frac{1 - \pi}{1 - \pi^*} h(\x).
\end{equation}
Since \(\pi \ne \pi^*\) and \(h(\x) \ne h^*(\x)\), the pair \((\pi, h(\x))\) is not uniquely determined, rendering the model unidentifiable and thus unsuitable for statistical inference without further constraints.

A common solution is to impose a parametric structure on the density ratio between \(g(\x)\) and \(h(\x)\) (e.g., \citealt{lancaster_case-control_1996}).
The \textit{exponential tilting model}, widely adopted in the literature, assumes the form:
\begin{equation}
	\label{eq:etm}
	\log \left\{ \frac{g(\x)}{h(\x)} \right\} = \alpha + \x\trans \bb.
\end{equation}
The exponential tilting model in \eqref{eq:etm} is closely connected with using logistic regression for case-control studies. In fact, under the assumption of a logistic regression $\bbP(y = 1 \mid \x) = \exp(\alpha^* + \x{\trans}{\boldsymbol{\beta}})/\{1 + \exp(\alpha^* + \x{\trans}{\boldsymbol{\beta}})\}$,
the density ratio satisfies
$
{f_1{(\x)}}/{f_0(\x)} = \exp(\alpha + \x{\trans}\boldsymbol{\beta}),
$
where $f_1$ and $f_0$ denote the conditional density of $\X$ given $y = 1$ (case group) and $y = 0$ (control group), respectively, and $\alpha = \alpha^* + \log\{(1 - \pi)/\pi\}$ with $\pi$ denoting the marginal probability
of $y = 1$ in the population. Under the exponential tilting model, a wide range of tasks, including parameter estimation, classification, and inference, become feasible. 

The exponential tilting model addresses the identifiability problem by imposing a linear form on the log-density ratio~\citep[see][for a recent proof]{liu2024positive}. However, the model's reliance on a parametric linear structure also constitutes its main limitation. If the true log-density ratio is not linear in \(\x\), the model may be misspecified, potentially leading to poor performance. Our numerical studies later in this paper show that model misspecification can severely affect estimation and prediction.

This situation presents a fundamental dilemma: Allowing the negative class distribution \(h(\x)\) to be completely nonparametric leads to non-identifiability, as shown in \eqref{eq:not-identifiable}.
Conversely, enforcing a strong parametric assumption (e.g., linearity) ensures identifiability but may result in severe model misspecification. We are thus motivated to ask:

\begin{center}
	\textit{Can we construct a model that is both identifiable and more flexible than the parametric exponential tilting model?}
\end{center}

\subsection{Literature review}
PU data have been studied by both the machine learning and statistics communities, and the existing work can be broadly divided into two major lines.
The first line focuses on training classifiers: \citet{elkan_learning_2008} proposed a probabilistic formulation for PU learning, which inspired a large body of classifier-oriented methods (e.g., \citealt{plessis2014analysis, kiryo_positive-unlabeled_2017, kato_learning_2018, chen_variational_2020}).
The second line emphasizes modeling and mixture-proportion estimation, including work on contaminated case-control and presence-only models (e.g., \citealt{lancaster_case-control_1996, ward2009presence, duan2019fast, song2019pulasso, liu2024positive}) and nonparametric or semi-parametric approaches to estimating mixture proportions (e.g., \citealt{blanchard2010semi, scott_rate_2015, garg_mixture_2021}).

Across these frameworks, researchers impose various assumptions to ensure identifiability.
Some works simply assume the mixture proportion is known (\citealt{kato_learning_2018}); others rely on post-hoc identifiability conditions, meaning that although the estimation algorithm itself does not include the identifiability constraint, the assumption is invoked afterward to guarantee that the resulting estimator is well-defined (e.g., the anchor-set assumption in \citealt{garg_mixture_2021}; the maximum-canonical assumption in \citealt{scott_rate_2015}).
Lastly, most if not all works in the statistical literature adopt the parametric modeling assumptions described in (\ref{eq:etm}) (\citealt{lancaster_case-control_1996, ward2009presence, duan2019fast, song2019pulasso, liu2024positive}).

\blue{Among these studies, \citet{liu2024positive} is most closely related to our work. They develop a double exponential tilting model (DETM) that relaxes the SCAR assumption by allowing the labeled and unlabeled positive samples to follow different distributions, while retaining parametric linear specifications for the density ratios. Our work takes a different but complementary direction. We retain the SCAR assumption but replace the linear density-ratio model with a flexible additive structure. The main contribution of the present work is therefore to establish identifiability, estimation, and inference for the mixture proportion and the unknown additive components under this more flexible density-ratio model.}

\subsection{Our contributions}
Under the SCAR assumption in \eqref{eq:pu-dist}, we address the question raised in Section~\ref{sub:exponential_tilting_model} by proposing the \textit{generalized additive exponential tilting model}. 
At a high level, the key idea is to replace the restrictive linear structure \(\eta(\x) \equiv \x\trans \bb\) in~\eqref{eq:etm} with an additive form \(\eta(\x) = \sum_{j=1}^{p} u_j(x_j)\), where \(x_j\) is the \(j\)th component of \(\x\), and each \(u_j(\cdot)\) is an unspecified function.  
This generalization parallels the way a generalized additive model extends classical linear regression.

Our main contributions are as follows:
\begin{enumerate}
	\item We propose the generalized additive exponential tilting model for PU data and provide conditions under which the model is identifiable.
	\item We develop a sieve-based estimation procedure for the parameters \((\pi, u_1, \dots, u_p)\) under the empirical likelihood framework, along with a practical EM-type algorithm for implementations.
	\item We establish the asymptotic properties of the proposed estimator, including consistency, asymptotic normality for \(\pi\), and weak convergence for the nonparametric functions \(u_j(\cdot)\), for \(j = 1, \dots, p\). These theoretical guarantees enable us to develop tools to carry out statistical inference for $\pi$ and $u_j$'s. 
\end{enumerate}

The remainder of the paper is organized as follows.  
Section~\ref{sec:gaet} introduces the generalized additive exponential tilting model and discusses its identifiability and likelihood function.  
Section~\ref{sec:sieve-estimator} presents the nonparametric estimators. Section~\ref{sec:algorithm} describes the computational algorithm for calculating the estimates. Section~\ref{sec:theory} develops the asymptotic theory. Section~\ref{sec:numerical} contains numerical results.  
Section~\ref{sec:conclude} concludes the paper. Some additional implementation details, empirical results,
%including applications to two semi-synthetic datasets, 
and all theoretical proofs are relegated to the Supplementary Material. 

%\vspace{-4mm}

\section{Generalized additive exponential tilting model}
\label{sec:gaet}

Built on the formulation in~\eqref{eq:pu-dist}, the {generalized additive exponential tilting} (GAET) model specifies the density ratio between the positive and negative features distributions as
\begin{equation}
	\label{eq:gaet}
	\log\left\{\frac{g(\x)}{h(\x)}\right\} = \alpha + \sum_{j=1}^{p} u_j(x_j),
\end{equation}
where each \( u_j(\cdot) \) is a nonparametric function satisfying the centering constraint \( \int u_j(x_j) \, \de x_j = 0 \), and \( x_j \) denotes the \( j \)th component of the \( p \)-dimensional feature vector \( \x \). The scalar parameter \( \alpha \) acts as a normalizing constant to ensure that the density ratio integrates to one under the negative class distribution \( h(\x) \), i.e.,
\begin{equation}
	\label{eq:normalizing-constant}
	\int \exp\left\{\alpha + \sum_{j=1}^{p} u_j(x_j) \right\} h(\x) \, \de \x = 1.  \end{equation}
In model \eqref{eq:gaet}, it is implicitly assumed that 
the supports of $h$ and $g$ overlap. 
Compared with the traditional exponential tilting model in \eqref{eq:etm}, the GAET model expands the model class by allowing flexible, additive, nonparametric components.
However, a key question is under what conditions the GAET model is identifiable.

While the complete identifiability conditions involve technical subtleties, we find that the GAET model is generally identifiable when \( p \geq 2 \). The precise conditions are given in the following theorem.

\begin{theorem}\label{thm:indentify-1}
	Suppose \( p \geq 2 \) and \( \pi \in (0,1) \). 
	Assume that 
	the densities \( g(\cdot) \) and \( h(\cdot) \) are both positive over a compact set \( \mathcal{T} = \mathcal{T}_1 \times \cdots \times \mathcal{T}_p \), where \( \mathcal{T}_j \) is a compact subset of the domain of \( u_j(\cdot) \), for \( j = 1, \dots, p \). 
	Assume \eqref{eq:gaet} holds on $\mathcal T$ and \( \int_{\mathcal T_j} u_j(x_j) \, \de x_j = 0 \).  
	If there exist two distinct indices \( j, k \in \{1, \dots, p\} \) such that, for each $l\in\{j,k\}$,  $\mathcal T_l$ contains at least two points and \( u_l(\cdot) \) is non-constant on $\mathcal T_l$, then the model defined by~\eqref{eq:pu-dist} and~\eqref{eq:gaet} uniquely determines the parameters \( (\pi, \alpha, u_1, \dots, u_p) \).
\end{theorem}

\begin{remark}
	The condition \( p \geq 2 \) is necessary, as the GAET model is not identifiable when \( p = 1 \) and a counterexample can be constructed using \eqref{eq:not-identifiable}. For any \(\pi^* \in (0, \pi)\), define $h^*$ by \eqref{eq:not-identifiable} and let $\alpha^*=c^{-1}\int_0^1 \log\left\{{g(x_1)}/{h^*(x_1)}\right\}\de x_1$, where $c$ denotes the Lebesgue measure of $\mathcal T_1$.  
	We then define $u_1^*(x_1)= \log\{g(x_1)/h^*(x_1)\}-\alpha^*$ to ensure \( \int_0^1 u^*_1(x_1) \, \de x_1 = 0 \). Consequently, both $(\pi,\alpha,u_1,g,h)$ and $(\pi^*,\alpha^*,u^*_1,g,h^*)$ satisfy~\eqref{eq:pu-dist} and~\eqref{eq:gaet}.
	
	Although we usually consider continuous $\X$ in model \eqref{eq:gaet}, 
	Theorem \ref{thm:indentify-1} can be extended to include discrete random variables, with a centering constraint of the form $\sum_{x_j} u_j(x_j)=0$ for each discrete $X_j$. 
	Using the same arguments as in its proof, we can show that Theorem \ref{thm:indentify-1} 
	still holds when at least one of $X_j$ and $X_k$ is discrete, 
	provided additionally that $1<\# \mathcal T_l<\infty$ for each discrete covariate $X_l$, where $l\in\{j,k\}$ and $\# \mathcal T_j$ denotes the cardinality of $\mathcal T_j$. 
\end{remark}

\section{Estimation via empirical likelihood and sieve methods}
\label{sec:sieve-estimator}

Recall that we use \( n_0 \) to denote the number of labeled (positive) instances and \( n \) the total sample size.  
Define the density ratio by \( \omega(\x) = g(\x)/h(\x) \).  
Given the observed samples \( \{\x_1, \dots, \x_{n_0}\} \) and \( \{\x_{n_0+1}, \dots, \x_n\} \), the likelihood function under~\eqref{eq:pu-dist} is 
\begin{equation}
	\label{eq:likelihood1}
	\begin{split}
		L_n(\pi, g, h) =\; &\prod_{i=1}^{n_0} g(\x_i) \prod_{j=n_0+1}^n \left\{ \pi g(\x_j) + (1 - \pi) h(\x_j) \right\} \\
		=\; &\left\{ \prod_{i=1}^n h(\x_i) \right\}
		\left\{ \prod_{i=1}^{n_0} \omega(\x_i) \right\}
		\prod_{j=n_0+1}^n \left\{ \pi \omega(\x_j) + (1 - \pi) \right\}.
	\end{split}
\end{equation}
Under the GAET model, the density ratio takes the form
$
\omega(\x) = \exp\left\{ \alpha + \sum_{j=1}^p u_j(x_j) \right\}$, 
and so the likelihood in \eqref{eq:likelihood1}  can be written as
\begin{equation}
	\label{eq:mle}
	L_n(\pi, g, h)=L_n(\pi, \alpha, u_1, \dots, u_p, h).
\end{equation}
Our primary goal is to maximize the likelihood with respect to \( \pi \in (0,1) \), $\alpha\in\mathbb R$, and the additive components \( u_j(\cdot) \) for $j=1,\dots, p$. Since an accurate estimate of \( h(\cdot) \) is not of direct interest, the negative class density \( h(\cdot) \) is treated as a nuisance parameter.

We have established the connection between the logistic regression and the exponential tilting model after model \eqref{eq:etm}. Similar results hold for the generalized additive logistic regression and our GAET model at \eqref{eq:gaet}. 
However, it should be noted that, in logistic regression, the binary outcome is available for each subject. Therefore, the likelihood for the observed outcomes conditional on the features can be written directly, depending only on the additive form in the log odds. In contrast, in PU data, we cannot access labels in the unlabeled group. 
To leverage the information of the (positively) labeled group, we instead consider the likelihood function of $\x$ rather than the likelihood function of $y$ conditional on $\x$. Because of the unknown labels for features in the unlabeled group, 
the likelihood function for the unlabeled group is represented in terms of a mixture model. Therefore, the likelihood function depends on $\omega(\x)$, $h(\x)$, and $\pi$, as specified in \eqref{eq:likelihood1}. Consequently, a direct application of the numerical tools for fitting generalized additive models cannot address our estimation problem.

There are two main challenges in the maximization of the likelihood $L_n(\pi,\alpha,u_1,\dots,u_p,h)$ in \eqref{eq:mle}: the first is addressing the unknown base density \( h(\cdot) \), which is a \( p \)-dimensional probability density function; the second is addressing the nonparametric functions \( u_j(\cdot) \), for \( j = 1, \dots, p \).
Our solution to these two challenges is to combine the empirical likelihood and sieve methods.

%\vspace{-7pt}

\subsection{Profiling out \texorpdfstring{$h(\x)$}{h(x)} via empirical likelihood}

To deal with the nuisance parameter $h(\cdot)$, one may substitute an estimate of $h$ into \eqref{eq:mle}. However, 
estimating the negative class density \( h(\cdot) \) is a challenging task, particularly in moderate-to-high dimensions.
Importantly, specifying a parametric model for \( h(\cdot) \) is generally infeasible due to its complexity and lack of structural assumptions.
%\sout{Moreover, since our primary focus is on the parameters \( (\pi, \alpha, u_1, \dots, u_p) \), accurate recovery of \( h(\cdot) \) is not of direct interest.  
	%For these reasons, we treat \( h(\cdot) \) as a nuisance function and}
Moreover, directly plugging an estimate of \( h(\cdot) \) into the likelihood function can elevate difficulties in the theoretical development. For these reasons, we propose to profile out the nuisance $h(\cdot)$  using the empirical likelihood method. We can also choose to profile out $g(\cdot)$; this will lead to the same inference.

Specifically, we let $p_i=h(\x_i)$ for $i=1,\dots n$ and treat them as parameters in (\ref{eq:likelihood1}).
After taking the log function, the empirical log-likelihood function is 
\begin{align}
	\begin{split}\label{eq:ell}
		& \ell_n(\pi,\alpha,u_1,\dots,u_p,p_1,\dots,p_n)\\
		& =\sum_{i=1}^{n}\log(p_i)+\sum_{i=1}^{n_0}\log\left\{\omega(\x_i)\right\}+\sum_{i=n_0+1}^n\log\left\{\pi\omega(\x_i)+(1-\pi)\right\},
	\end{split}
\end{align}
where $(p_1,\dots,p_n)\in\calC_n(\alpha,u_1,\dots,u_p)$, which is defined as
\[\calC_n(\alpha,u_1,\dots,u_p)=
\left\{(p_1,\dots,p_n):p_i>0,i=1,\dots,n;\sum_{i=1}^np_i=1;\sum_{i=1}^np_i\omega(\x_i)=1\right\}.
\]
The constraints $p_i>0,i=1,\dots,n$ and $\sum_{i=1}^n p_i=1$ are due to the fact that $h(\x)$ is a probability density function; the constraint $\sum_{i=1}^n p_i\omega(\x_i)=1$ is from (\ref{eq:normalizing-constant}).

Using the method of Lagrange multipliers, we can find that the maximum value of $\ell_n$ is attained at $p_i=n^{-1}\left[1+\lambda\left\{\omega(\x_i)-1\right\}\right]^{-1}$ for $i=1,\dots,n$, where $\lambda$ is the Lagrangian multiplier.
The empirical log-likelihood $\ell_n(\lambda; \pi, \alpha, u_1, \ldots, u_p)$, after profiling out $p_i,i=1,\dots,n$ and removing a constant, becomes
\begin{equation}
	\label{eq:ell_no_pi}
	\begin{split}
		%\ell_n(\lambda; \pi, \alpha, u_1, \ldots, u_p) = & 
		-\sum_{i=1}^n \log[1 + \lambda\{\omega(\mathbf{x}_i) - 1\}] + \sum_{i=1}^{n_0} \log\{\omega(\mathbf{x}_i)\}+ \sum_{i=n_0+1}^n \log\{\pi \omega(\mathbf{x}_i) + (1 - \pi)\},
	\end{split}
\end{equation}
where the Lagrange multiplier $\lambda$ satisfies
\begin{equation}
	\label{eq:lagrange-constraint}
	\sum_{i=1}^n \frac{\omega(\x_i)-1}{1+\lambda\left\{\omega(\x_i)-1\right\}}=0.
\end{equation}

Equation~(\ref{eq:lagrange-constraint}) implies that $\lambda$ is an implicit function of $\alpha$ and $u_j(\cdot)$, $j=1,\dots,p$.
The following proposition provides the value of $\lambda$ that maximizes $\ell_n$ defined in \eqref{eq:ell_no_pi}. 

%\vspace{-3.5mm}

\begin{proposition}
	\label{prop:lambda-max}
	When $\ell_n$ attains its maximal value, the Lagrangian multiplier is $\widetilde{\lambda}=(n_0+n_1\pi)/n$, where $n_1=n-n_0$.
\end{proposition}
By inserting $\lambda=\widetilde{\lambda}$ into (\ref{eq:ell_no_pi}), the profile log-likelihood for $(\pi,\alpha,u_1,\dots,u_p)$, which is free of $h(\cdot)$, is
\begin{equation}
	\label{eq:profile-final}
	\begin{split}
		\ell_n(\pi,\alpha,u_1,\dots,u_p)=&\sum_{i=1}^{n_0}\log\left[\frac{n_0\exp\left\{\alpha+\eta(\x_i)-\tau\right\}}{n_1\pi+(n_0+n_1\pi)\exp\left\{\alpha+\eta(\x_i)-\tau\right\}}\right]\\
		&+\sum_{i=n_0+1}^n\log\left[\frac{n_1\pi+n_1\pi\exp\left\{\alpha+\eta(\x_i)-\tau\right\}}{n_1\pi+(n_0+n_1\pi)\exp\left\{\alpha+\eta(\x_i)-\tau\right\}}\right],
	\end{split}
\end{equation}
where $\tau\equiv\log\left\{(1-\pi)/\pi\right\}$ and $\eta(\x_i)=\sum_{j=1}^p u_j(x_{ij})$.
Here $\x_i$ denotes the features of the $i$th instance, and $x_{ij}$ is the $j$th component of $\x_{i}$.

\subsection{Sieve approximation of the additive components}

In the previous subsection, we profile out the nuisance parameter \( h(\cdot) \) using an empirical likelihood approach. The remaining task is to estimate $\pi$, $\alpha$, and the additive component \( \eta(\x) = \sum_{j=1}^p u_j(x_j) \). Up to this point, we have referred to the functions \( u_j(\cdot) \), for \( j = 1, \dots, p \), as ``nonparametric'' in an informal sense, without specifying their mathematical form. In this subsection, we formalize these components using the sieve method, which provides a finite-dimensional approximation to the infinite-dimensional function space and enables estimation.

Specifically, we approximate each \( u_j(\cdot) \) by a linear combination of spline basis functions. To construct the spline basis, we first assume, without loss of generality, that the support of \( \x \) is the unit cube \( [0,1]^p \). Let \( C^d([0,1]) \) denote the class of functions with continuous derivatives up to order \( d \) on the interval \( [0,1] \).
We begin by partitioning the interval \( [0,1] \) using \( K_n \) interior knots:
%\begin{equation*}
$\Delta = \left\{ 0 = \xi_0 < \xi_1 < \cdots < \xi_{K_n} < \xi_{K_n+1} = 1 \right\}.$
%\end{equation*}
For a fixed integer \( m \geq 2 \), we define the spline function class of order \( m \) as
\begin{multline*}
	\calS_n = \left\{ f \in C^{m-2}([0,1]) : 
	f \text{ is a polynomial of degree at most } m - 1 \right. \\
	\left. \text{on each interval } [\xi_j, \xi_{j+1}], ~ 0 \leq j \leq K_n \right\}.
\end{multline*}
Each function \( f \in \calS_n \) admits a B-spline basis representation of the form
\begin{equation} \label{eq-Bspline-approximation}
	f(x) = \sum_{k=1}^{K_n + m} \theta_k N_k^{[m]}(x), \quad x \in [0,1],
\end{equation}
where \( \{N_k^{[m]}(\cdot)\}_{k=1}^{K_n + m} \) are normalized B-spline basis functions of order \( m \), and \( \{\theta_k\}_{k=1}^{K_n + m} \) are the associated coefficients (\citealt[Chapter~4]{schumaker2007spline}).
The basis functions satisfy \( 0 \leq N_k^{[m]}(x) \leq 1 \) for all \( x \in [0,1] \) and \( k = 1, \dots, K_n + m \).

Based on the B-spline representation in~\eqref{eq-Bspline-approximation}, we approximate each component \( u_j(\cdot) \) in~\eqref{eq:profile-final} by a function \( \nu_j(\cdot) \in \calS_n \) of the form
$
u_j(x_j) \approx \nu_j(x_j) = \sum_{k=1}^{K_n + m} \theta_{jk} N_k^{[m]}(x_j)$, for $j = 1, \dots, p$.
Then we define the sieve space as 
\begin{equation}
	\label{eq:likelihood-sieve-space}
	\begin{aligned}
		\calF_n(m, K_n) = \Big\{ & (\pi, \alpha, \nu_1, \dots, \nu_p) : \pi \in [\minpi, 1 - \minpi],\; |\alpha| \leq M, \\
		& \nu_j(x_j) = \sum_{k=1}^{K_n + m} \theta_{jk} N_k^{[m]}(x_j), \\
		& \max_{1 \leq k \leq K_n + m}|\theta_{jk}| \leq q_j, \quad \int_0^1 \nu_j(x) \, \de x = 0,\quad j = 1, \dots, p \Big\}.
	\end{aligned}
\end{equation}
Finally, the sieve estimators of \( (\pi, \alpha, u_1, \dots, u_p) \) are defined as the maximizer of the profile log-likelihood:
\begin{equation}
	\label{eq:maximizer}
	(\widehat{\pi}_n, \widehat{\alpha}_n, \widehat{u}_{1,n}, \dots, \widehat{u}_{p,n})
	= \argmax_{(\pi, \alpha, \nu_1, \dots, \nu_p) \in \calF_n(m, K_n)} 
	\ell_n(\pi, \alpha, \nu_1, \dots, \nu_p),
\end{equation}
where \( \ell_n(\cdot) \) is the profile log-likelihood defined in~\eqref{eq:profile-final}.

\begin{remark} \label{rmk:sieve}
	In the sieve space~\eqref{eq:likelihood-sieve-space}, we restrict \( |\alpha| \leq M \) and \( \pi \in [\minpi, 1 - \minpi] \) for some known constants \( M \) and \( \minpi > 0 \).  
	The centering condition \( \int_0^1 \nu_j(x) \, \de x = 0 \) aligns with the constraint imposed in~\eqref{eq:gaet}. 
	Based on the fact that $\sum_{k = 1}^{K_n + m} N_k^{[m]}(x) = 1$ for any $x \in [0, 1]$, the assumption $\max_{1 \leq k \leq K_n + m}|\theta_{jk}| \leq q_j$ implies that $\|\nu_j\|_{\infty} \leq q_j$. 
	The upper bounds \( \|\nu_j\|_\infty \leq q_j \) for $j=1,\dots,p$ are used to guarantee compactness of the sieve space. As suggested in \citet{zeng2005likelihood}, a commonly used choice for the sieve parameters is \( (m, K_n) = (k + 2, C n^\beta) \) for some constant \( C > 0 \) while the choice of \( \beta \) and \( k \) will be discussed in Section~\ref{sec:theory}.
\end{remark}

\section{Algorithm}
\label{sec:algorithm} 

We develop an algorithm for parameter estimation under the GAET model.
Solving the optimization problem \eqref{eq:maximizer} directly is challenging, since $\ell_n(\pi,\alpha,\nu_1,\dots,\nu_p)$ is nonconvex. \cite{ward2009presence} noted that algorithms based on direct maximization of the observed likelihood under model \eqref{eq:etm} may fail to converge, especially when $\pi$ is unknown. 
They therefore suggested an expectation–maximization (EM) algorithm \citep{dempster1977maximum}. 
In this section, we first establish the full likelihood and then develop an EM-type algorithm, in which each iteration solves a convex optimization problem. 
Moreover, we show that our algorithm provides a reasonable solution to \eqref{eq:maximizer}. 
We refer to the proposed algorithm as an EM-type algorithm because the M-step maximizes a profile log-likelihood.

\subsection{The full likelihood and the E-step}
Suppose the labels $y_i$ are observed for all subjects $i=1,\dots,n$. 
By \eqref{eq:pu-dist} and \eqref{eq:gaet}, the full log-likelihood for $(\x, y)$ is
\begin{equation*}
	\begin{split}
		&\sum_{i=1}^{n_0} \log \{g(\x_i)\}+\sum_{i=n_0+1}^{n} \left[ y_i\log\{\pi g(\x_i)\}+(1-y_i)\log\{(1-\pi)h(\x_i)\} \right]\\
		%=\,&\sum_{i=1}^{n_0} \log \{h(\x_i)\omega(\x_i)\}+\sum_{i=n_0+1}^{n} \left[ y_i\log\{\pi h(\x_i)\omega(\x_i)\}+(1-y_i)\log\{(1-\pi)h(\x_i)\} \right]\\
		=\,&  \sum_{i=1}^n \log\{h(\x_i)\}+\sum_{i=1}^{n_0}\log \{\omega(\x_i)\} +\sum_{i=n_0+1}^n \left[y_i\log \{\pi\omega(\x_i)\}+(1-y_i)\log(1-\pi)\right] ,
	\end{split}
\end{equation*}
where $\omega(\x_i)$ is defined in Section 2. 
%$\omega(\x_i)=\exp\{\alpha+\eta(\x_i)\}$ as defined in \eqref{eq:mle} and $\eta(\x_i)=\sum_{j=1}^p u_j(x_{ij})$ as defined in \eqref{eq:profile-final}.  

The E-step computes the conditional expectation of the full log-likelihood given observed data and the current parameters. 
Denote by $\pi^{[r]}$, $\alpha^{[r]}$ and $u_j^{[r]}(\cdot)$ the estimates after the $r$\textsuperscript{th} iteration, where $r$ is a non-negative integer with $r=0$ corresponding to the initial values.  
Let $\eta^{[r]}(\x)=\sum_{j = 1}^p u_j^{[r]}(x_j)$ for $\x = (x_1, \ldots, x_p)\trans$. 
For $i=n_0+1,\dots,n$, let $y_i^{[r+1]}$ denote the expectation of the label $y_i$ conditional on the observed data given $\pi^{[r]}$, $\alpha^{[r]}$ and $\eta^{[r]}(\cdot)$. 
By Bayes' rule and \eqref{eq:gaet}, we have 
\begin{equation}\label{eq:E-step}
	y_i^{[r+1]}=\frac{\pi^{[r]} \exp\{\alpha^{[r]}+\eta^{[r]}(\x_i)\}}{\pi^{[r]} \exp\{\alpha^{[r]}+\eta^{[r]}(\x_i)\}+1-\pi^{[r]}},\quad i=n_0+1,\dots,n.
\end{equation}
The E-step is then performed by substituting $y_i$ in the full log-likelihood with $y_i^{[r+1]}$. 

\vspace{-5mm}

\subsection{The M-step and the monotonicity}

The M-step maximizes the conditional expectation obtained in the E-step with respect to $\pi,\alpha$ and $\eta$ by profiling $h(\x_i)$'s via empirical likelihood. 
Recall that $p_i=h(\x_i)$, $i=1,\dots,n$. 
The conditional expectation of the full empirical log-likelihood after substituting $p_i$ is
\begin{align}
	\begin{split}
		\label{eq:Mstep00}
		%\label{eq:ell-full}
		Q^{[r+1]}(\pi,\alpha,\eta,\p)&=\sum_{i=1}^{n}\log(p_i)+\sum_{i=1}^{n_0}\log\left\{\omega(\x_i)\right\}\\
		&+\sum_{i=n_0+1}^n \left[y_i^{[r+1]}\log \{\pi\omega(\x_i)\}+(1-y^{[r+1]}_i)\log(1-\pi)\right], 
	\end{split}
\end{align}
for $r=0,1,\dots$, where $\p=(p_1,\dots,p_n){\trans}\in\calC_n(\alpha,u_1,\dots,u_p)$ defined in \eqref{eq:ell}.
Then, the M-step is performed by maximizing $Q^{[r+1]}(\pi,\alpha,\eta,\p)$ subject to $\p\in\calC_n(\alpha,u_1,\dots,u_p)$ and $(\pi, \alpha, \eta) \in \calF_n(m, K_n)$ given in \eqref{eq:likelihood-sieve-space}.

Observe that $Q^{[r+1]}(\pi,\alpha,\eta,\p)=Q_1^{[r+1]}(\alpha,\eta,\p)+Q_2^{[r+1]}(\pi)$, 
where $Q_1^{[r+1]}(\alpha,\eta,\p)=\sum_{i=1}^n \log p_i+\sum_{i=1}^n y^{[r+1]}_i\{\alpha+\eta(\x_i)\}$ 
and 
$Q_2^{[r+1]}(\pi)=\sum_{i=n_0+1}^n \{y^{[r+1]}_i\log \pi+(1-y^{[r+1]}_i)\log(1-\pi)\}$. 
Hence, we update $\pi$ by maximizing $Q_2^{[r+1]}(\pi)$ only: 
\begin{equation}\label{eq:M-step-1}
	\pi^{[r+1]}=\argmax_{\pi\in [0, 1]}\,Q_2^{[r+1]}(\pi) =\frac1{n_1}\sum_{i=n_0+1}^n y^{[r+1]}_i.
\end{equation}

%Next we consider optimization of $Q_1^{[r+1]}(\alpha,\eta,\p)$ with respect to $\alpha$, $\eta$ and $\p$. 
Compared with \citet{liu2024positive}, the key contribution of our algorithm resides in the optimization of $Q_1^{[r+1]}(\alpha,\eta,\p)$, in which we obtain a nonlinear function optimizer $\eta$ under a sieve framework via regression splines, whereas \citet{liu2024positive} only requires fitting a generalized linear model to yield a linear $\eta$.
Similarly to Proposition \ref{prop:lambda-max}, we can show that the optimal $p_i$ takes the form
%\begin{equation*}
$p_i=n^{-1}[1+\lambda^{[r+1]}\{\omega(\x_i)-1\}]^{-1}$, 
%\end{equation*}
where $\lambda^{[r+1]}=n^{-1}(n_0+\sum_{i=n_0+1}^n y^{[r+1]}_i)$, $i=1,\dots,n$. 
Substituting these $p_i$'s into $Q_1^{[r+1]}(\alpha,\eta,\p)$ and 
discarding irrelevant terms, we arrive at the following objective function to be maximized: 
\begin{equation}
	\label{eq:M-step-2}
	\sum_{i=1}^n y^{[r+1]}_i\{\alpha_*+\eta(\x_i)\}-\sum_{i=1}^n \log[1+\exp\{\alpha_*+\eta(\x_i)\}],
\end{equation}
where $\alpha_*=\alpha+c^{[r+1]}$ with $c^{[r+1]}=\log(n_0/n_1+\pi^{[r+1]})-\log(1-\pi^{[r+1]})$. 
Note that the form of \eqref{eq:M-step-2} coincides with the log-likelihood function of the generalized additive logistic model without penalty, %\citep[Chapter 9]{hastie2009elements}, 
except that working responses $y^{[r+1]}_i$ are continuous and lie within $[0,1]$. 
This similarity allows us to employ some 
existing algorithms developed for nonparametric logistic regressions. 
In our numerical implementation, the R function \texttt{gam} from the \texttt{mgcv} package \citep{wood2017generalized} is applied. 
The pseudo-code of the EM-type algorithm is presented in Algorithm \ref{algo-EM-1}. It is worth noting that, in our estimation procedure, except estimating the additive density ratio $\alpha + \eta(\x)$, the other steps are not needed when fitting the generalized additive logistic regression.

\begin{algorithm}
	\caption{EM-type algorithm for the GAET model.}\label{algo-EM-1}
	\begin{algorithmic}[1]
		\Require initial values $\pi^{[0]}$, $\alpha^{[0]}$, $u_j^{[0]}$, $j=1,\dots,p$.
		\For {$r= 0,1,\ldots,$ }
		\State E-step:  update $y_i^{[r+1]}$ by \eqref{eq:E-step}. 
		\State M-step:  update $\pi^{[r+1]}$, $\alpha_*^{[r+1]}$ and $u_j^{[r+1]}$ by \eqref{eq:M-step-1} and \eqref{eq:M-step-2}. 
		\State Shifting: $\alpha^{[r+1]} \leftarrow \alpha_*^{[r+1]}-c^{[r+1]}$ with  $c^{[r+1]}$ defined below   \eqref{eq:M-step-2}.
		\EndFor
		\State  \textbf{Output} $\pi^{[r_0]}$, $\alpha^{[r_0]}$ and $u_j^{[r_0]}$ for some $r_0$ when the stopping criteria are met.
	\end{algorithmic}
\end{algorithm}

%The following proposition shows that the values of the observed profile log-likelihood function evaluated at the EM-type outputs are non-decreasing with respect to iterations. 
\vspace{-5mm}
\begin{proposition}
	\label{prop:em}
	Let $(\pi^{[r]},\alpha^{[r]},\eta^{[r]})$ be the estimates obtained by Algorithm \ref{algo-EM-1} at the $r$\textsuperscript{th} iteration, where $\eta^{[r]}(\x)=\sum_{j = 1}^p u_j^{[r]}(x_j)$. Then for any $r \geq 0$, we have 
	$$\ell_n(\pi^{[r]},\alpha^{[r]},\eta^{[r]})\leq \ell_n(\pi^{[r+1]},\alpha^{[r+1]},\eta^{[r+1]}),$$ 
	where $\ell_n(\pi,\alpha,\eta) \equiv \ell_n(\pi,\alpha,u_1,\dots,u_p)$ is given by \eqref{eq:profile-final}. 
\end{proposition}

Proposition \ref{prop:em} shows that the observed profile log-likelihood, evaluated at the outputs of the proposed algorithm, is non-decreasing across iterations. 
This result extends \citet[Proposition~4]{ward2009presence}, which assumes that $\pi$ is known and $\eta(\cdot)$ is linear in $\x$. 
From \eqref{eq:profile-final}, it follows that $\ell_n(\pi,\alpha,\eta)\leq 0$. 
Proposition \ref{prop:em} then guarantees that Algorithm \ref{algo-EM-1} eventually converges. Furthermore, Proposition \ref{prop:em} supports our claim that the EM-type algorithm provides a reasonable solution to \eqref{eq:maximizer}.

\section{Theoretical results}
\label{sec:theory}

With the sieve-based profile likelihood estimators in~\eqref{eq:maximizer}, we now turn to the theoretical properties of the proposed estimators. Specifically, we establish consistency and asymptotic normality for the estimators of the mixture proportion \( \pi \), the normalization constant \( \alpha \), and the additive components \( u_j(\cdot) \), for \( j = 1, \dots, p \).

%Letting $n_0$ and $n_1$ be the sample sizes of labeled and unlabeled data and $n=n_0+n_1$ be the total sample size, 
We assume that the ratio of  the sample sizes of labeled and unlabeled data, $\rho=n_0/n_1$, is a constant.
This assumption can be relaxed to $\lim_{n\to\infty}n_0/n_1=\rho$, but we hold the ratio as a constant to simplify the presentation.
We denote the true values of $\pi$ and $\alpha$ as $\pi_0$ and $\alpha_0$, and use $\eta_0(\cdot)$ to denote the true additive component $\eta(\x)=\sum_{i=1}^p u_j(x_j)$.
As in \eqref{eq:maximizer}, the estimators of $\pi$ and $\alpha$ are denoted as $\widehat{\pi}_n$ and $\widehat{\alpha}_n$, respectively.
Let $\widehat{\eta}_n(\cdot)$ denote the sieve estimator of $\eta(\cdot)$.  

We need the following assumptions to establish theoretical results.
\begin{assumption} \label{ass:parameter}
	There exist known constants $M$ and $\minpi \in (0, 1/2)$ such that $\alpha_0$ and $\pi_0$ satisfy $|\alpha_0| \leq M$
	and $\pi_0 \in [\minpi, 1 - \minpi]$, respectively. 
\end{assumption}
\begin{assumption} \label{ass:support}
	The support of $\X$ is $[0, 1]^p$, and there exists a constant $C_0 > 0$ such that $C_0^{-1} \leq h(\x) \leq C_0$ for any $\x \in [0, 1]^p$. 
\end{assumption}
\begin{assumption} \label{ass:Holder}
	For $1 \leq j \leq p$, $u_j \in C^{k}([0, 1])$ for some $0 < k \leq m$ and satisfies $\int_0^1 u_j(x) \de x = 0$, where $m$ is the order of the spline functions.
\end{assumption}
\begin{assumption} \label{ass:Knorder}
	The number of interior knots $K_n$ satisfies $K_n \rightarrow \infty$ and $K_n^{1/2} \log K_n = o(\sqrt{n})$ as $n \rightarrow \infty$. 
\end{assumption}

\begin{remark}[On the Assumptions]
	Assumption~\ref{ass:parameter} imposes lower and upper bounds on the mixture proportion \( \pi \), ensuring that relevant functions in the estimation procedure remain Lipschitz continuous with respect to \( \pi \). This type of condition is frequently adopted in semiparametric models for the purpose of asymptotic analysis (e.g., \citealt{murphy1995asymptotic, chen2013estimating}).
	We have already taken this assumption into consideration when defining $\calF_n(m, K_n)$ in \eqref{eq:likelihood-sieve-space}. 
	%commonly referred to as the {overlap assumption} in the causal inference literature (e.g., \citealt{yang2018asymptotic}), and it is frequently adopted in semiparametric models for the purpose of asymptotic analysis (e.g., \citealt{murphy1995asymptotic, chen2013estimating}).
	
	Assumptions~\ref{ass:support} and~\ref{ass:Holder} are standard in the literature on nonparametric estimation using regression splines within exponential family models.
	These conditions ensure identifiability and control the approximation error of the sieve basis functions; for related discussions, see \citet{stone1986dimensionality} and \citet{xue2010consistent}.

	Assumption~\ref{ass:Knorder} regulates the growth rate of the number of interior knots \( K_n \). This condition reflects the classic bias-variance trade-off in nonparametric estimation: increasing \( K_n \) expands the flexibility of the sieve space, thus reducing approximation bias, but at the cost of increased estimation variance. The stated rate \( K_n^{1/2} \log K_n = o(\sqrt{n}) \) ensures that the trade-off is balanced to guarantee consistency and asymptotic normality of the estimator. 
	%Further discussion on the choice of \( K_n \) in spline-based sieve estimators can be found in \citet{xiao2019asymptotic}.
\end{remark}

Then, we establish consistency results in the following theorems.

\begin{theorem} \label{thm:consistency}
	Under Assumptions \ref{ass:parameter} to \ref{ass:Knorder}, $\widehat{\alpha}_n$, $\widehat{\pi}_n$, and $\widehat{\eta}_n$ are consistent estimators of $\alpha_0$, $\pi_0$ and $\eta_0$, respectively.
\end{theorem}
Furthermore, we can obtain the convergence rate of the estimators in the following theorem.

\begin{theorem} \label{thm:rate}
	Under Assumptions \ref{ass:parameter} to \ref{ass:Knorder}, the following results hold:

	\begin{align*}
		|\widehat{\pi}_n - \pi_0|^2 & \leq O_p\left(\frac{1}{K_n^{2k}}\right) + o_p\left(\frac{1}{\sqrt{n}}\right),~|\widehat{\alpha}_n - \alpha_0|^2 \leq O_p\left(\frac{1}{K_n^{2k}}\right) + o_p\left(\frac{1}{\sqrt{n}}\right), \\
		\text{and}\qquad \qquad \quad  & \\
		\left\|\widehat{\eta}_n - \eta_0\right\|^2_{L_2(\bbP)} & \leq O_p\left(\frac{1}{K_n^{2k}}\right) + o_p\left(\frac{1}{\sqrt{n}}\right).
	\end{align*}
\end{theorem}

Then we develop the asymptotic distributions of the estimators. 
%We consider the asymptotic distribution of the estimator of the logarithm of the density ratio first. 
Let $\calG_{k} \subset C^k([0, 1])$ consist of all functions defined in [0, 1] whose $k$th derivative is bounded by 1, where $k$ is specified in Assumption \ref{ass:Holder}.
Let $\mathcal{O}_{\alpha}$ and $\mathcal{O}_{\pi}$ be the unit interval in $\mathbb{R}$.
Following \cite{murphy1995asymptotic} and \cite{chen2013estimating}, we treat $\{\widehat{\eta}_n(\x) - \eta_0(\x), \widehat{\alpha}_n - \alpha_0, \widehat{\pi}_n - \pi_0\}$ as an element in $l^{\infty}(\oplus_{j = 1}^p \calG_{k} \times \mathcal{O}_{\alpha} \times \mathcal{O}_{\pi})$, which is the class of bounded functions defined on $\oplus_{j = 1}^p \calG_{k} \times \mathcal{O}_{\alpha} \times \mathcal{O}_{\pi}$.
In particular, for any $\{\sum_{j = 1}^p g_j(x_j), b_{\alpha}, b_{\pi}\} \in \oplus_{j = 1}^p \calG_k \times \mathcal{O}_{\alpha} \times \mathcal{O}_{\pi}$ where $g_j \in \calG_k$ for $j = 1, \ldots, p$, the evaluation functional is
\begin{equation} \label{eq-linfty}
	\sum_{j = 1}^p \int_0^1 g_j(x_j) \{\widehat{u}_j(x_j) - u_{j0}(x_j)\} \de x_j +  b_{\alpha}(\widehat{\alpha}_n - \alpha_0) + b_{\pi}(\widehat{\pi}_n - \pi_0). 
\end{equation}
%\syf{\sout{where $\widehat{\eta}_n(\x) = \sum_{j = 1}^p \widehat{u}_j(x_j)$.}}
The following theorem gives the asymptotic distribution of $\{\widehat{\eta}_n(\x) - \eta_0(\x), \widehat{\alpha}_n - \alpha_0, \widehat{\pi}_n - \pi_0\}$, where $\eta_0(\x) = \sum_{j = 1}^p u_{j0}(x_j)$ and $\widehat{\eta}_n(\x) = \sum_{j = 1}^p \widehat{u}_j(x_j)$.
\begin{theorem} \label{thm:asympnormality-np}
	Under Assumptions \ref{ass:parameter} to \ref{ass:Knorder} and $\sqrt{n} = o(K_n^{2k})$, $\sqrt{n}\{\widehat{\eta}_n(\x) - \eta_0(\x), \widehat{\alpha}_n - \alpha_0, \widehat{\pi}_n - \pi_0\}$ converges in distribution to a mean-zero Gaussian process in the metric space $l^{\infty}(\oplus_{j = 1}^p \calG_k \times \mathcal{O}_{\alpha} \times \mathcal{O}_{\pi})$ with the evaluation functional defined in \eqref{eq-linfty}. 
\end{theorem}
As a direct consequence of Theorem \ref{thm:asympnormality-np}, we establish the asymptotic distribution of $\sqrt{n} (\widehat{\pi}_n - \pi_0)$, which allows us to construct valid confidence intervals for $\pi_0$. 
\begin{corollary} \label{thm:asympnormality}
	Under Assumptions \ref{ass:parameter} to \ref{ass:Knorder} and $\sqrt{n} = o(K_n^{2k})$, $\sqrt{n}(\widehat{\pi}_n - \pi_0)$ converges to a normal distribution as $n \rightarrow \infty$. 
\end{corollary}

These results provide rigorous justification for the validity of our estimation procedure.
Our theoretical contributions are novel in several key aspects:
\begin{itemize}
	\item Most existing work on PU data assumes that the mixture proportion \( \pi \) is known \citep[e.g.,][]{song2019pulasso, song2020convex}. 
	While \citet{ward2009presence} considered the case of unknown \( \pi \), they did not provide any theoretical or empirical analysis of its estimation.
	One exception is \citet{ramaswamy_mixture_2016}, but they only proved consistency.
	In contrast, our framework allows \( \pi \) to be unknown and provides both consistency and asymptotic normality results for its estimator.
	To the best of our knowledge, this is the first work to derive the asymptotic distribution of the estimator for \( \pi \) under the SCAR assumption in the PU setting.
	\item Our theoretical analysis is also novel in its use of a combined empirical likelihood and sieve estimation framework.
	While both techniques have been studied independently, their integration is underexplored in the literature, and we are not aware of any prior results establishing asymptotic theory under this combination. Empirical likelihood has been used in case-control studies \citep{qin1998inferences} and PU learning \citep{liu2024positive} under the parametric exponential tilting model, and theoretical results are developed using techniques in \cite{qin1994empirical}.
	%Based on the theories of applying empirical likelihood to estimating equations developed in \cite{qin1994empirical}, one can  establish theoretical properties of the estimated parameters in the density ratio.
	In contrast, such technical tools have not been developed for a nonparametric additive density ratio that is estimated through a sieve method in the literature.
	In this work, we make use of semiparametric tools to fill in this theoretical gap.
\end{itemize}

%Furthermore, we propose a bootstrap confidence interval for $\pi$ based on Corollary \ref{thm:asympnormality} in Section \ref{sec:bootstrap}.  

\section{Numerical studies}
\label{sec:numerical}

%The implementation code and detailed instructions for accessing the datasets used in this article are available at 
%\url{https://github.com/syf6250/GAET_PU}.

\subsection{Practical considerations}\label{sec:prac}

\noindent{\textbf{Roughness penalty and stopping criteria}}
\,\,
To better balance bias and variance in estimating these additive components, we %use a large number of B-spline basis functions in $\calF_n(m, K_n)$, and 
incorporate a roughness penalty term %\citep{green1993nonparametric,xiao2019asymptotic} 
into the original objective function:
$
\tilde{Q}^{[r + 1]}(\pi,\alpha,\eta,\p) = Q^{[r+1]}(\pi,\alpha,\eta,\p) -P_{\kappa_n}(\eta),
$
where $Q^{[r+1]}(\pi,\alpha,\eta,\p)$ is defined in \eqref{eq:Mstep00} and $P_{\kappa_n}(\eta)=n\kappa_n \sum_{j=1}^p\int_0^1 \ddot{\nu}_j^2(t) \de t$ with
$\kappa_n$ being a tuning parameter and $\ddot{\nu}_j(\cdot)$ being the second derivative of ${\nu}_j(\cdot)$. 
%This strategy is often referred to as penalized splines \citep{green1993nonparametric, xiao2019asymptotic}. 
To maximize \eqref{eq:M-step-2} with penalty term $P_{\kappa_n}(\eta)$ in the M-step, we can still employ the R function \texttt{gam} from the \texttt{mgcv} package.

To implement Algorithm \ref{algo-EM-1}, we need to specify the initial values, which are critical because the EM-type algorithm may converge to distinct stationary points of \eqref{eq:profile-final} under different initializations. 
In our experiment, we generate 20 random initial values and compute the observed profile log-likelihood function \eqref{eq:profile-final} evaluated on the 20 corresponding estimates. 
The output with the largest likelihood value is chosen as the final estimates. 
The methods for generating initial values are discussed in Section S.2.1 of the Supplementary Material.

We stop the iterations in Algorithm \ref{algo-EM-1} if the following two conditions are satisfied simultaneously: (i) the increment of $\ell_n(\pi^{[r]},\alpha^{[r]},\eta^{[r]})-P_{\kappa_n}(\eta^{[r]})$ in two consecutive iterations is less than a pre-specified threshold,  
and (ii) the relative error for the estimates of $\pi$, i.e., $|\pi^{[r]}-\pi^{[r-1]}|/\pi^{[r-1]}$, is smaller than a threshold. 
In the numerical studies, we set $10^{-4}$ as the threshold for both conditions.

\noindent{\textbf{Tuning parameter selection}}
\,\,
To implement the algorithm, determining the values of $K_n$, $m$ and $\kappa_n$ is needed. 
Our strategy is to choose the tuning parameter $\kappa_n$ while fixing the number of basis functions, $K_n+m$, at a moderate level \citep[Section 5.5]{ruppert2003semiparametric}. 
In the subsequent sections of this paper, we follow the package default \citep{wood2017generalized} and set $m=4$ and $K_n+m=10$. 
Additional simulation work shows that the choice of $K_n$ has little impact on the estimation of $\pi$ and functions $u_j$, $j=1,\dots,p$. 
%A data-driven AIC method is then adopted to determine ${\kappa_n}$. The details can be found in Section \ref{sec:tuning} of the Supplementary Material. 

We develop an AIC-based method to determine the tuning parameter ${\kappa_n}$ in \eqref{eq:M-step-2}. When solving the generalized additive model \eqref{eq:M-step-2} with a penalty term, the \texttt{gam} function computes and returns the effective degrees of freedom \citep{wood2020inference}. 
Denote by $df_{\kappa_n}$ the effective degrees of freedom for the final iteration of the EM-type algorithm with tuning parameter ${\kappa_n}$. 
Then we seek the optimal ${\kappa_n}$ such that the resulting \(-2\ell_n(\hat\pi_{\kappa_n}, \hat\alpha_{\kappa_n}, \hat\eta_{\kappa_n})+2\, df_{\kappa_n}\) 
is minimized, 
where $(\hat\pi_{\kappa_n}, \hat\alpha_{\kappa_n}, \hat\eta_{\kappa_n})$ are the estimates given ${\kappa_n}$, and  $\ell_n(\pi,\alpha,\eta) \equiv \ell_n(\pi,\alpha,u_1,\dots,u_p)$ is given by \eqref{eq:profile-final}. 
See Section S.2.2 of the Supplementary Material for a sensitivity analysis of our method relative to the choice of $\kappa_n$ for the simulation studies in Section \ref{sec:simulation}.

\subsection{Bootstrap confidence interval and hypothesis testing}\label{sec:bootstrapCI}
Note that the asymptotic variance of  $\hat{\pi}_n$ has no closed form (see the proof of Theorem \ref{thm:asympnormality-np} in the Supplementary Material for more details).
To address this issue, the nonparametric bootstrap method 
% \citep[Section 4.1.5] {shao2012jackknife} %\citep{efron1979bootstrap}  
is applied to construct confidence intervals for $\pi$.
More specifically, we resample from the labeled data and the unlabeled data separately, and implement Algorithm \ref{algo-EM-1} based on the resampled data. 
Repeat the procedure $B$ times, and let $\hat\pi_j^*$ denote the estimate of $\pi$ for the $j$th replicate, $j=1,\dots, B$. 
% Let $\hat\pi$ represent the estimate using the original data. 
A $(1 - \alpha)$-confidence interval of $\pi$ is then given by 
$[2\hat\pi_n-\hat q_B (1 - \tfrac{\alpha}{2}),\,  2\hat\pi_n-\hat q_B ( \tfrac{\alpha}{2})]$,
where $\hat q_B(s)$ represents the lower $s$-quantile of the sequence $\{\hat\pi_j^*: j=1,\dots, B\}$. 
%We can find level-90\% confidence interval similarly. 
In the following simulation studies, we additionally apply the logit transformation to the estimated $\pi$ and the true $\pi$ to check the empirical coverage rates.

To test the significance of any nonparametric additive component, say $u_1$, we consider the projected vector $\widehat{\bf b}=(\hat{b}_1,\dots, \hat{b}_{L})\trans$, where $\hat{b}_l =\int_0^1 \hat u_1(x_1)\phi_l(x_1) \de x_1$ with $\phi_l$ being the $l$\textsuperscript{th} Fourier basis function over $[0,1]$ for $l = 1, \ldots, L$. 
Throughout this paper, we set $L=11$ in numerical experiments.  
By Theorem \ref{thm:asympnormality-np}, $\sqrt n\, \widehat{\bf b}$ converges to a mean zero normal distribution under the null hypothesis $u_1(x_1) = 0$ for $x_1 \in [0, 1]$. Based on the continuous mapping theorem, we consider the test statistic $n\widehat{\bf b}\trans \widehat{\bf b}$. 
The null hypothesis is rejected if its value exceeds a critical value with the nominal significance level of 0.05. 
Using the aforementioned resampling scheme, we obtain the corresponding estimates ${\bf b}^{[a]}$, $a=1,\dots,B$. 
The critical value is the 95\% quantile of 
$\{n({\bf b}^{[a]}-\widehat{\bf b})\trans ({\bf b}^{[a]}-\widehat{\bf b}):a=1,\dots,B\}.$ 
%A relatively large $L$ is recommended, and we choose $L = 11$ in our numerical implementations. 

\subsection{Simulation studies}\label{sec:simulation}

\paragraph{\textbf{Data generation process}}

In the simulation studies, we set $p=5$ and  $\pi_0$ as 0.4.
%Let $(n_0,n)\in \{(1250, 1500), (2500, 3000), (3750, 4500), (6250,7500)\}$. 
Let $n\in\{1500,3000,4500,7500\}$ and set $n_0= 5n/6$. 
The datasets are generated as follows:  
\begin{enumerate}
	\item Generate a binomially distributed variable $n_*$ with success probability $\pi_0$ and $n-n_0$ trials.
	\item Independently draw each component of $\X$ from $U(0,1)$. For each realized $\x$, generate a binary label $y$ from the logistic model $\Pr(Y=1\mid \X=\x)=\text{expit}\{m(\x)\}$ for a pre-specified function $m(\x)$ (details given later). 
	If $y=1$, then $\x$ is classified as Group 1; otherwise, $\x$ is classified as Group 0.
	\item Repeat Step 2 until the numbers of Group 1 and Group 0 are at least $n_0+n_*$ and $n-n_0-n_*$, respectively.  
	The labeled sample is collected as the first $n_0$ observations from Group 1; the unlabeled sample consists of the next $n_*$ observations from Group 1 together with $n-n_0-n_*$ observations from Group 0. 
\end{enumerate}

By \eqref{eq:pu-dist} and Bayes' rule, we can show that $m(\x)$ and the log-density ratio differ only by an additive constant. Therefore, rather than working directly with the density ratio $g(\x)/h(\x)$, we control the data-generating mechanism through $m(\x)$. The choices of $m(\x)$ considered in our simulations are listed below: 
\begin{enumerate}
	\item A linear function $m(\x)=-16+6\sum_{j=1}^5 x_j$. %Approximately, $\mathbb P(Y=1)=0.409$.
	\item A quadratic function $m(\x)=-14+24\sum_{j=1}^5 (x_j-0.3)^2$. %Approximately, $\mathbb P(Y=1)=0.509$.
	\item A general nonlinear function $m(\x)=-13+6x_1+24(x_2-0.3)^2+1/(x_3+0.1)-5\cos(5x_4)+2\exp(4x_5-2)$.
\end{enumerate}
%\noindent \textit{Setting 1 (linear)}: Let $m(\x)=-16+6\sum_{j=1}^5 x_j$. Approximately, $\mathbb P(Y=1)=0.409$. 
% The candidates for $\kappa_n$ are $\{10^{a}:a=-1,0,\dots,4\}$. 

%\noindent \textit{Setting 2 (quadratic)}: Let $m(\x)=-14+24\sum_{j=1}^5 (x_j-0.3)^2$. Approximately, $\mathbb P(Y=1)=0.509$. 
% The candidates for $\kappa_n$ are $\{10^{a}:a=-6,-5,\dots,-1\}$.  

%\noindent \textit{Setting 3 (general)}: 
%Let $m(\x)=-13+6x_1+24(x_2-0.3)^2+1/(x_3+0.1)-5\cos(5x_4)+2\exp(4x_5-2)$. Approximately, $\mathbb P(Y=1)=0.470$. 
% The candidates for $\kappa_n$ are $\{10^{a}:a=-6,-5,\dots,-1\}$.  

\paragraph{\textbf{Simulation results}}

To evaluate the accuracy of $\pi_0$ estimation, we compute the bias, standard error, and mean squared error of $\hat\pi_n$ over 2000 replicates.
Moreover, for $i=n_0+1,\dots,n$, we define the posterior classifier by $\widehat y_i=\mathbbm 1(y_i^{[r_0]}>0.5)$, where $\mathbbm 1(\cdot)$ is the indicator function and $y_i^{[r_0]}$ is the output of the final E-step in \eqref{eq:E-step}. 
To assess classification performance, we compute the false positive rate, false negative rate, 
\[\textup{FP} =\frac{\#\{i\in I_u:y_i=0,\widehat y_i=1\}}{\#\{i\in I_u:y_i=0\}},\;
\textup{FN}=\frac{\#\{i\in I_u:y_i=1,\widehat y_i=0\}}{\#\{i\in I_u:y_i=1\}},\]
and overall classification error  
$\textup{Err}=(n-n_0)^{-1}\, \#\{i\in I_u:y_i\neq\widehat y_i\}$, 
%\[\textup{Err}=\frac{\#\{i\in I_u:y_i\neq\widehat y_i\}}{\# I_u},\]
where $I_u=\{n_0+1,\dots,n\}$ and $\#A$ represents the cardinality of set $A$. 

We compare the proposed method with the parametric exponential tilting model in~\eqref{eq:etm} (denoted as ``linear'').
The comparison results are reported in Table \ref{tab:pest}. 
As a benchmark, Table~\ref{tab:pest} also reports the performance of the Bayes classifier (with the largest sample size), assuming $\pi_0$ and the true density ratio are known.

\begin{table}[htbp]
	\centering
	\caption{Bias, standard error (se), and mean squared error (mse) of the estimated $\pi$ ($\times10^2$) for 2000 replicates; together with FP, FN and Err of classifiers.}
	%\vspace*{5mm}
	\setlength\tabcolsep{8.5pt}%{6pt}
	\begin{tabular}{ccccccccc}
		\toprule
		Setting & $n$   & method & bias  & se    & mse   & FP    & FN    & Err \\
		\midrule
		1     & \multirow{2}[1]{*}{1500} & GAET  & -0.245  & 5.098  & 0.261  & 0.099  & 0.185  & 0.132  \\
		&       & linear & -0.170  & 5.079  & 0.258  & 0.099  & 0.183  & 0.132  \\
		& \multirow{2}[0]{*}{3000} & GAET  & -0.195  & 3.616  & 0.131  & 0.097  & 0.177  & 0.129  \\
		&       & linear & -0.168  & 3.610  & 0.131  & 0.098  & 0.176  & 0.129  \\
		& \multirow{2}[0]{*}{4500} & GAET  & -0.024  & 2.857  & 0.082  & 0.098  & 0.172  & 0.127  \\
		&       & linear & -0.012  & 2.857  & 0.082  & 0.098  & 0.172  & 0.127  \\
		& \multirow{2}[0]{*}{7500} & GAET  & -0.050  & 2.177  & 0.047  & 0.097  & 0.171  & 0.127  \\
		&       & linear & -0.050  & 2.176  & 0.047  & 0.097  & 0.171  & 0.127  \\
		&       & Bayes & -     & -     & -     & 0.097  & 0.168  & 0.125  \\
		\addlinespace[3pt]%\midrule
		2     & \multirow{2}[1]{*}{1500} & GAET  & -1.801  & 4.494  & 0.234  & 0.045  & 0.145  & 0.084  \\
		&       & linear & -5.206  & 7.117  & 0.778  & 0.083  & 0.374  & 0.197  \\
		& \multirow{2}[0]{*}{3000} & GAET  & -1.680  & 2.866  & 0.110  & 0.041  & 0.133  & 0.077  \\
		&       & linear & -4.959  & 4.987  & 0.495  & 0.079  & 0.359  & 0.190  \\
		& \multirow{2}[0]{*}{4500} & GAET  & -1.300  & 2.276  & 0.069  & 0.042  & 0.124  & 0.075  \\
		&       & linear & -5.001  & 3.970  & 0.408  & 0.076  & 0.356  & 0.187  \\
		& \multirow{2}[0]{*}{7500} & GAET  & -0.869  & 1.579  & 0.032  & 0.044  & 0.115  & 0.072  \\
		&       & linear & -4.909  & 3.064  & 0.335  & 0.076  & 0.352  & 0.186  \\
		&       & Bayes & -     & -     & -     & 0.050  & 0.100  & 0.070  \\
		\addlinespace[3pt]%\midrule
		3     & \multirow{2}[1]{*}{1500} & GAET  & -2.023  & 4.701  & 0.262  & 0.053  & 0.165  & 0.097  \\
		&       & linear & 8.864  & 8.151  & 1.450  & 0.251  & 0.181  & 0.222  \\
		& \multirow{2}[0]{*}{3000} & GAET  & -1.601  & 3.038  & 0.118  & 0.050  & 0.146  & 0.088  \\
		&       & linear & 8.655  & 5.908  & 1.098  & 0.248  & 0.176  & 0.219  \\
		& \multirow{2}[0]{*}{4500} & GAET  & -1.204  & 2.448  & 0.074  & 0.052  & 0.137  & 0.086  \\
		&       & linear & 8.926  & 4.847  & 1.032  & 0.252  & 0.169  & 0.218  \\
		& \multirow{2}[0]{*}{7500} & GAET  & -0.884  & 1.845  & 0.042  & 0.053  & 0.129  & 0.083  \\
		&       & linear & 8.935  & 3.629  & 0.930  & 0.251  & 0.166  & 0.217  \\
		&       & Bayes & -     & -     & -     & 0.058  & 0.112  & 0.080  \\
		\bottomrule
	\end{tabular}%
	\label{tab:pest}%
\end{table}%

In Setting 1 where the true density ratio model is linear, the proposed method performs similarly to the linear method; both methods give good classification results, which are only slightly worse than those of the oracle Bayes classifier. In contrast, when the density ratio model is nonlinear, as in Settings 2--3, our method remarkably outperforms the linear method in terms of both estimating $\pi$ and classification.
Furthermore, as the sample size $n$ grows, the misclassification error of our method decreases and becomes close to that of the optimal Bayes classifier.
%\blue{We perform a sensitivity analysis for the performance of our method relative to the choice of $\kappa_n$ under these settings; see Section \ref{sec:tuning_study} of the Supplementary Material.}

\begin{figure}[htbp]
	\centering
	\includegraphics[width=0.9\linewidth]{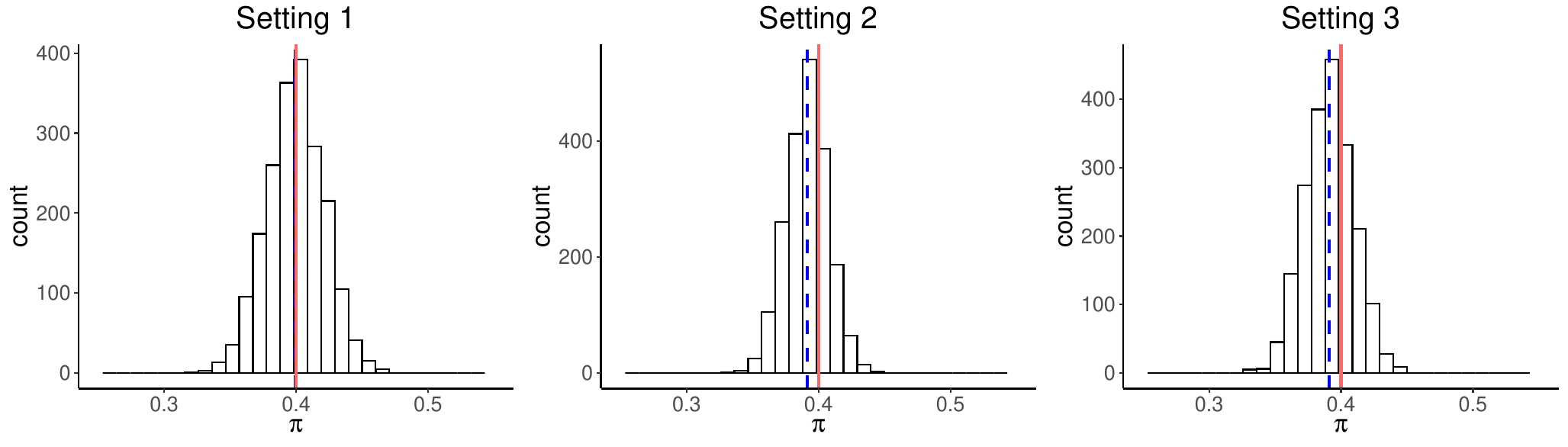}
	\caption{Histograms of estimated $\pi$ of 2000 replicates under various settings with $n=7500$. Red solid vertical line: $\pi_0=0.4$; blue dotted vertical line: the average of 2000 estimates.}
	\label{fig:piest}
\end{figure}

Next we investigate the performance of statistical inference for $\pi_0$ based on the bootstrap confidence intervals. 
Figure \ref{fig:piest} displays frequency histograms of 2000 estimates of $\pi_0$ under the three settings with $n=7500$. %Further normality testing results are deferred to Section \ref{sec:normal_test} of the Supplementary Material. 
The approximately bell-shaped distributions confirm the conclusion in Corollary \ref{thm:asympnormality}, with additional supporting results provided in Section S.2.3 of the Supplementary Material. Table \ref{tab:iest} summarizes the empirical coverage probabilities for the 95\% bootstrap confidence interval of $\pi_0$ over the $2000$ replicates. The number of bootstrap resamples for each simulated dataset is set as $500$. Table \ref{tab:iest} indicates that
the empirical coverage probability becomes close to the nominal level as the sample size $n$ increases. 
Specifically, if the true density ratio follows the exponential tilting model, bootstrap confidence intervals achieve near-nominal coverage even for small $n$, whereas in nonlinear settings a larger sample size ($n \ge 3000$) is needed for satisfactory coverage.

\begin{table}[htbp]
	\centering
	\caption{Empirical coverage probabilities for 95\% bootstrap confidence intervals of $\pi_0$ over 2000 replicates.}
	\setlength\tabcolsep{12pt}
	%\vspace*{5mm}
	%\renewcommand\arraystretch{1}
	\begin{tabular}{cccc}
		\toprule
		$n$   & Setting 1 & Setting 2 & Setting 3 \\
		\midrule
		1500  & 0.959  & 0.726  & 0.609  \\
		3000  & 0.952  & 0.934  & 0.941  \\
		4500  & 0.948  & 0.963  & 0.955  \\
		7500  & 0.957  & 0.953  & 0.957  \\
		\bottomrule
	\end{tabular}%
	\label{tab:iest}%
\end{table}%

\begin{figure}[htbp]
	\centering
	\includegraphics[width=0.75\linewidth]{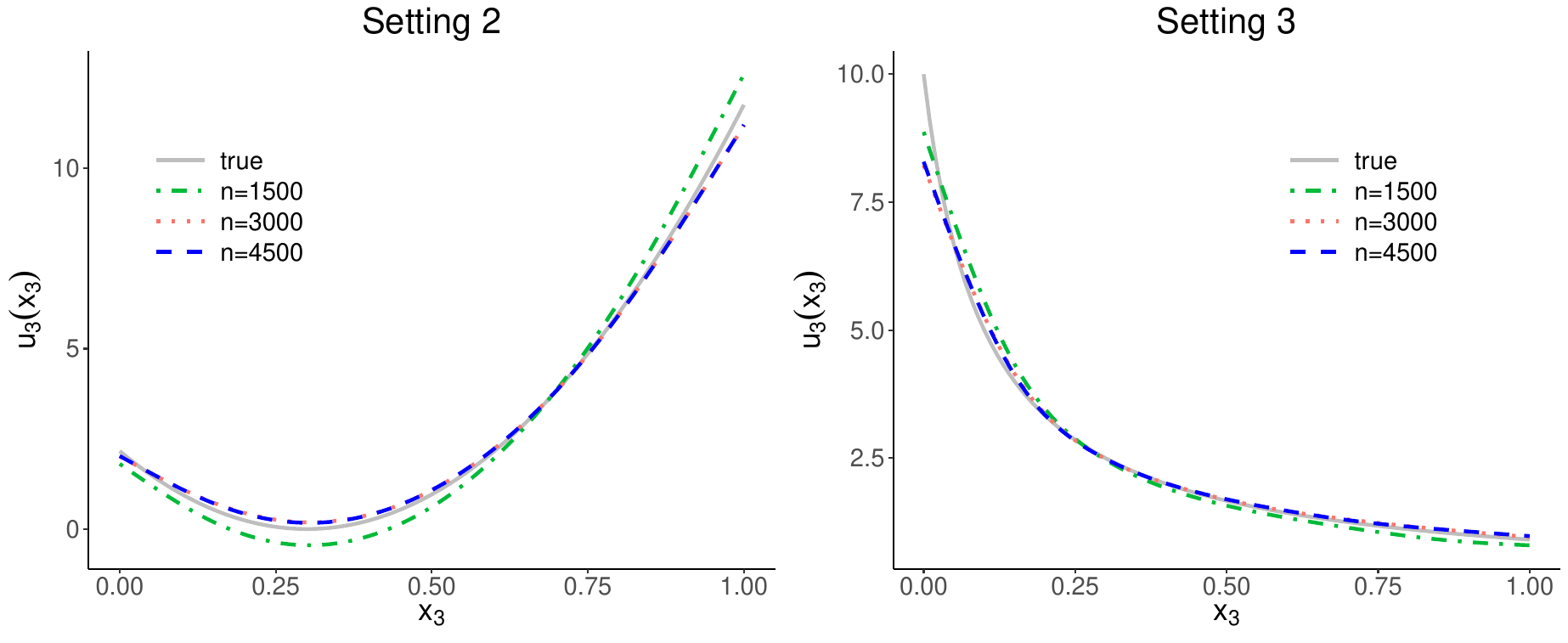}
	\caption{Mean of estimated $u_3(\cdot)$ of 2000 replicates under various scenarios. Dash-dotted green: $n=1500$; red dotted: $n=3000$; blue dash: $n=4500$; grey solid: true functions.}
	\label{fig:fun2est}
\end{figure}

We further investigate the estimation and inference of the additive components.
Figure \ref{fig:fun2est} shows that, under Settings 2 or 3, 
the estimate of $u_3(x_3)$ averaged over 2000 replicates aligns well with the true underlying function. 
% \syf{\sout{In Setting 2, the estimate aligns well with the true function $u_3(\cdot)$ when the sample size is moderately large.
		% Similar patterns can be found in the right panel of Figure \ref{fig:fun2est} for Setting 3.}} 
In Setting 1, the averaged estimate is nearly indistinguishable from the truth for all sample sizes considered, and is therefore omitted. 
%In Setting 3, the performance improves as $n$ grows, similar to Setting 2. However, since the true function does not belong to the sieve class, the estimated curve only approximates the true curve, even when $n$ is 3000. The mean estimated curves are almost perfect for all sample sizes under Setting 1, and thus omitted. 
To test the significance of $u_1(\cdot)$, we replace $m(\x)$ in Settings 1 and 2 by  $m(\x)=-16+6\zeta x_1+6\sum_{j=2}^5 x_j$ and $m(\x)=-14+24\zeta (x_1-0.3)^2+24\sum_{j=2}^5 (x_j-0.3)^2$, respectively, and we refer to these two modified settings as Settings $1^*$ and $2^*$. 
Here $\zeta\in [0,1]$ is a positive constant that reflects the magnitude of $u_1(\cdot)$. 
Under each setting with a specific $\zeta$, the estimate $\widehat u_1$ is obtained by the proposed method. 
We apply the bootstrap-based inference method developed in Section \ref{sec:bootstrapCI} and report the empirical sizes and powers in Table \ref{tab:test-u1}. The empirical size is close to the nominal level when the sample size is large, and the empirical power converges to 1 as $\zeta$ or $n$ increases. 

\begin{table}[htbp]
	\centering
	\caption{Empirical sizes and powers over 2000 replicates for testing $H_0\colon u_1 = 0$ using a bootstrap test.}
	\setlength\tabcolsep{12pt}%{10pt}
	%\vspace*{5mm}
	%\renewcommand\arraystretch{1}
	\begin{tabular}{rcccccc}
		\toprule
		\multicolumn{1}{c}{Setting} & $n$ & $\zeta=0$ & 0.05  & 0.10  & 0.15  & 0.20  \\
		\midrule
		\multicolumn{1}{c}{$1^*$} & 1500  & 0.025  & 0.065  & 0.566  & 0.961  & 0.994  \\
		& 3000  & 0.033  & 0.243  & 0.999  & 1.000  & 1.000 \\
		& 4500  & 0.042  & 0.497  & 1.000  & 1.000  & 1.000 \\
		& 7500  & 0.043  & 0.990  & 1.000  & 1.000  & 1.000 \\
		\midrule
		\multicolumn{1}{c}{$2^*$} & 1500  & 0.018  & 0.046  & 0.230  & 0.593  & 0.736  \\
		& 3000  & 0.025  & 0.145  & 0.803  & 0.979  & 0.997  \\
		& 4500  & 0.031  & 0.257  & 0.981  & 1.000  & 1.000  \\
		& 7500  & 0.043  & 0.539  & 1.000  & 1.000  & 1.000 \\
		\bottomrule
	\end{tabular}%
	\label{tab:test-u1}%
\end{table}%

\blue{Additional simulations with a smaller value of $\pi_0$ ($\pi_0 = 0.1$), presented in Section S.2.4 of the Supplementary Material, show overall patterns similar to those under $\pi_0 = 0.4$, with improved coverage probabilities and powers in some settings.}

\subsection{Real data analyses}
\label{sec:realdata}
%We apply the proposed method to study the Special Eurobarometer survey 402 on undeclared work in the European Union \citep{arezzo2024estimating}. 
%Undeclared work refers to paid activities that are legal but concealed from public authorities.  
%Estimating the true size of undeclared work is important because misreporting in surveys can substantially understate its prevalence, potentially misleading public authorities about the real extent of undeclared work and undermining the design of effective policy measures. 

We first apply the proposed method to the survey dataset described in Section~\ref{sub:PUdata}.  
The data contain participants' answers to the question: ``Apart from a regular employment, have you yourself carried out any undeclared paid activities in the last 12 months?"  
Approximately 4.6\% of respondents answered “yes” and are treated as reliably positive; the remaining responses (“no”, refusal to answer, or missing) are treated as unlabeled. 
Due to potential dishonest answers, the true prevalence of undeclared work is expected to exceed the naive rate of 4.6\% \citep{arezzo2024estimating}.

Following \citet{arezzo2024estimating}, the variables of interest include gender, occupation status, financial problems, detection risks, age, and tax morale. Tax morale is a discrete index computed from responses to seven questions on attitudes toward fraudulent behaviors. Age is treated as continuous. All remaining variables (excluding age and tax morale) are categorical and dummy encoded into eight binary indicators. 
These indicators and tax morale are included as additional linear components in our GAET model. After removing subjects with missing covariates, we obtain $n_0=1145$ labeled positives and $n=24231$ total observations. 
%The dataset can be found at \url{https://search.gesis.org/research_data/ZA5688}. 

Using the proposed method under the GAET model \eqref{eq:gaet}, the point estimate of $\pi$ is 0.082, with a 95\% bootstrap confidence interval of $[0.040,0.124]$. The implied prevalence of undeclared work is then
$\hat\pi\cdot n_1/n + n_0/n = 0.125$.
For comparison, the linear method under \eqref{eq:etm} yields $\hat\pi=0.097$, resulting in an estimated prevalence of 0.139. Both estimates are substantially larger than the naive prevalence 0.046. 
We also test the significance of the additive component of age using the procedure in Section \ref{sec:bootstrapCI}. The resulting $p$-value is below 0.01, indicating a statistically significant effect of age. 
The estimated nonlinear function of age, shown in Section S.2.5 of the Supplementary Material, is broadly aligned with the linear fit in overall trend, while the nonlinear specification provides a more flexible and potentially more accurate characterization of the age effect. 
%captures clear curvature (e.g., a midlife plateau and a steeper decline at older ages), providing 

Our second real data application is on the species presence-only ecological data. 
Specifically, we analyze an anonymized tree species (swi28) in Switzerland, extracted from the “SWI” dataset in the R package \texttt{disdat} \citep{elith2020presence}. 
The resulting data comprise a set of locations with recorded presences of swi28 and associated environmental covariates, together with a background sample of locations measured on the same covariates but with unknown presence status.  
In total, the dataset includes $n_0=5528$ presence records and $n_1=10000$ background (unlabeled) records.

For each location, 13 environmental predictors are available, summarizing vegetation cover, soil conditions, climate, and topography. 
We fit the GAET model using seven continuous covariates, including 
growing degree-days above $0^\circ$C, soil nutrients index, annual precipitation, slope, potential yearly global radiation, site water balance, and topographic position, along with six discrete covariates: broadleaved continuous cover, coniferous continuous cover, an indicator of strictly calcareous bedrock, summer frost frequency, average temperature of coldest month, and the number of days with rainfall exceeding 1 mm. These discrete variables are again incorporated into the GAET model as linear components.

Our method yields $\hat{\pi}=0.124$ with a 95\% bootstrap confidence interval of $[0.106,0.142]$, while the linear method gives an estimate of $0.146$. 
The linear estimate falls outside our confidence interval, indicating a discrepancy between the two approaches and suggesting that the linear specification may be inadequate for modeling the density ratio, potentially inducing non-negligible bias in estimating  $\pi$.  
The estimated functions of continuous variables are reported in Section~S.2.5, which exhibit pronounced non-linear patterns across covariates. 
The $p$-values for all continuous predictors are below $0.01$, indicating that these variables are significant in the GAET model and are informative for distinguishing presence from absence for the tree species.

We also apply our method and the linear method to two semi-synthetic datasets, where the true $\pi$ is known. It turns out that our method outperforms the competitor in terms of estimating $\pi$ and prediction on the unlabeled samples; see Section S.2.7 of the Supplementary Material for more details.

\vspace{-5mm}

\section{Concluding remarks and discussions} % (fold)
\label{sec:conclude}

Our simulation studies and real-data analyses demonstrate that the GAET model 
matches the traditional parametric approach in approximately linear settings and substantially outperforms it for nonlinear density ratios, while maintaining solid theoretical guarantees. These results highlight the GAET model’s ability to balance modeling flexibility with identifiability, which is an enduring challenge in PU learning. Compared with the exponential tilting model, the GAET model is less likely to suffer from model misspecification because of its flexibility. Moreover, under mild conditions we establish identifiability for our model, which is crucial in PU data.
%Last but not least, we develop theoretical guarantees for our estimator. In particular, we establish asymptotic normality for $\hat \pi_n$ and $\hat{u}_j$'s, and statistical inference tools under this more general framework. This is a significant contribution in PU learning.}

By combining additive nonparametric modeling with empirical likelihood and sieve estimation, the proposed method delivers accurate estimation of mixture proportions, effective classification of unlabeled data, and valid statistical inference. This combination of flexibility, identifiability, and practical performance makes the GAET model a robust and broadly applicable tool for analyzing PU data. The proposed algorithm can be extended to deal with high dimensional predictors with the aid of the group LASSO; see Section S.3 of the Supplementary Material.

For the sieve space $\calF_n(m, K_n)$ defined in \eqref{eq:likelihood-sieve-space},
although the exact values of \( q_j, j=1,\dots,p, \) are not crucial in practice, such boundedness conditions are standard in sieve estimation to facilitate theoretical analysis. 
Following the recommendation of \citet{cheng2011semiparametric}, 
we impose finite bounds \( q_j \) for technical convenience, though consistency and asymptotic normality can still be established under diverging \( q_j \) using chaining arguments; see \citet{zeng2005likelihood} for details. We choose the same number of interior knots of B-spline basis functions, $K_n$, for these additive components $u_j$'s. In fact, $K_n$ can be chosen to be different for each component. Moreover, though we consider the same smoothness condition on these components in Assumption \ref{ass:Holder}, this restriction can be relaxed as well.

\blue{A limitation of the proposed methodology is its reliance on the SCAR assumption. In the survey example, this assumption may be questionable because individuals who report undeclared work may differ in their feature distributions from those who conceal it. In the ecological application, recorded presence locations may be preferentially collected from more accessible areas. Therefore, the real-data analyses should be interpreted under the SCAR working assumption.} 

\blue{As discussed in the Introduction, the proposed GAET model and the DETM of \citet{liu2024positive} relax different restrictions of the exponential tilting framework. 
Additional comparisons of two models in Section S.2.6 of the Supplementary Material suggest that suggest that the impact of relaxing SCAR to SAR can vary across applications. 
A natural extension would combine these two directions by imposing additive structures on both density ratios under the SAR setting:}
\[
\log\!\left\{\frac{g_1(\x)}{g_2(\x)}\right\}
= \alpha_1 + \sum_{i=1}^p \nu_i(x_i),\qquad
\log\!\left\{\frac{g_1(\x)}{h(\x)}\right\}
= \alpha_2 + \sum_{j=1}^p u_j(x_j).
\]
Nevertheless, extending the GAET framework to the SAR setting is nontrivial. 
For instance, the identifiability conditions developed in this paper are no longer sufficient under SAR. 
In addition, the resulting model introduces practical challenges such as label-switching issues ($h(\x)$ and $g_2(\x)$ are interchangeable) and potential numerical instability arising from the simultaneous estimation of multiple nonparametric density ratios. 
A complete treatment of the SAR--GAET model is therefore beyond the scope of the present paper and will be pursued in future work.

\section*{Data Availability.}
The data and code used to reproduce the results in this study are publicly available at
\url{https://github.com/syf6250/GAET_PU}.

\section*{Figure Legends}

\noindent\textbf{Figure 1.}
Histograms of the estimated $\pi$ across 2000 replicates under various settings with $n=7500$. The red solid vertical line indicates the true value, $\pi_0=0.4$, and the blue dotted vertical line indicates the average of the 2000 estimates.

\noindent\textbf{Figure 2.}
Mean estimated $u_3(\cdot)$ across 2000 replicates under various scenarios. Dash-dotted green: $n=1500$; red dotted: $n=3000$; blue dashed: $n=4500$; grey solid: true functions.

\bibliographystyle{apalike}
\bibliography{bibref}

\end{document}